\documentclass[12pt]{article}

\usepackage{amssymb,amsmath}

\newcommand{\be}{\begin{equation}}
\newcommand{\ee}{\end{equation}}
\newcommand{\Dlt}{\Delta}
\newcommand{\dlt}{\delta}

\newcommand{\ep}{\varepsilon}
\newcommand{\al}{\alpha}
\newcommand{\ra}{\rightarrow}
\newcommand{\sgm}{\sigma}

\newcommand{\gm}{\gamma}
\newcommand{\om}{\omega}

\newcommand{\Gm}{\Gamma}

\newcommand{\cH}{{\cal H}}

\newcommand{\rgl}{\rangle}
\newcommand{\lgl}{\langle}

\begin{document}

\begin{center}
{\Large{\bf  

          Equilibration of quasi-isolated quantum systems} \\ [5mm]

                             V.I. Yukalov } \\ [3mm]

{\it Bogolubov Laboratory of Theoretical Physics, \\
        Joint Institute for Nuclear Research, Dubna 141980, Russia}

\end{center}

\vskip 3cm

\begin{abstract}

The evolution of a quasi-isolated finite quantum system from a 
nonequilibrium initial state is considered. The condition of quasi-isolation 
allows for the description of the system dynamics on the general basis, 
without specifying the system details and for arbitrary initial conditions.
The influence of surrounding results in (at least partial) equilibration and
decoherence. The resulting equilibrium state bears information on initial 
conditions and is characterized by a representative ensemble. It is shown 
that the system average information with time does not increase. The partial 
equilibration and non-increase of average information explain the 
irreversibility of time. 

\end{abstract}

\vskip 2cm

{\bf Keywords}: Quasi-isolated quantum systems, Equilibration, Decoherence,
Average information, Irreversibility of time

\newpage

\section{Introduction}

The feasibility of producing various kinds of finite quantum systems has
sparked recently a strong interest in studying nonequilibrium properties of 
such objects. As examples of finite systems, it is possible to mention spin
assemblies, quantum dots and wells, and trapped atoms. The current studies
follow mainly two trends. The first direction, started yet by 
Schr\"{o}dinger [1,2], considers thermalization of a {\it non-isolated} 
system coupled to a giant thermal reservoir. Related to this is the case of 
local equilibration, when a finite part of a large system is considered, with 
the rest of the system playing the role of a giant reservoir. The second track, 
pioneered by von Neumann [3,4], explores relaxation of {\it isolated} quantum 
systems to quasi-stationary states characterized by ergodic averages. The 
relaxation of composite finite systems also pertains to this track [5-7]. There 
have appeared a number of papers on these topics of equilibration. Extensive 
list of literature can be found in the recent review articles [8-13].

The aim of this paper is to study the third way, when a finite quantum system
is {\it quasi-isolated} so that, though there exists some surrounding, but it 
is such that it does not disturb the system. In mathematical terms, this means 
that the matrix representation of all operators from the algebra of local 
observables is not influenced by the presence of the surrounding. It turns out 
that such quasi-isolated systems enjoy interesting specific properties, 
essentially differing them from both non-isolated as well as isolated systems.

\section{Quasi-isolated quantum systems}

Let us consider a system described by a Hamiltonian $H_A$ acting on a Hilbert 
space ${\cal H}_A$. And let there exist a surrounding characterized by a 
Hamiltonian $H_B$ on a Hilbert space ${\cal H}_B$. The total Hamiltonian is 
the sum
\be
\label{1}
 H_{AB} = H_A + H_B + H_{int} \;  ,
\ee
where $H_{int}$ is an interaction part. Hamiltonian (1) does not depend on time
and acts on the space
\be
\label{2}
 \cH_{AB} = \cH_A \bigotimes \cH_B \;  .
\ee

The system statistical ensemble is the pair $\{\cH_{AB},\hat{\rho}_{AB}(t)\}$
of space (2) and statistical operator
\be
\label{3}
 \hat\rho_{AB}(t) = \hat U_{AB}(t) \hat\rho_{AB} \hat U_{AB}^+(t) \;  ,
\ee
with the evolution operator
\be
\label{4}
 \hat U_{AB} = \exp ( - i H_{AB} t ) \;  .
\ee
Here $\hat\rho_{AB} \equiv \hat\rho_{AB}(0)$. The statistical operator 
is normalized, ${\rm Tr}_{AB}\hat\rho_{AB}(t)=1$, where the trace is 
over space (2).

The quantities of interest are the averages of the operators $\hat{A}$ 
from the algebra of local observables acting on the space ${\cal H}_A$. 
Such averages correspond to the measurable quantities
\be
\label{5}
 \lgl \; \hat A(t) \; \rgl \equiv 
{\rm Tr}_{AB} \hat\rho_{AB}(t) \hat A  = {\rm Tr}_A \hat\rho_A(t) \hat A \;  ,
\ee
where
\be
\label{6}
\hat\rho_A(t) \equiv {\rm Tr}_B \hat\rho_{AB}(t) \; .
\ee

The system, by definition, is quasi-isolated from the environment when the 
system Hamiltonian is conserved, such that
\be
\label{7}
 [ H_A , \; H_{AB} ] = 0 \;  .
\ee
As a consequence, we have
$$
[ H_A , \; H_B + H_{int} ] = [ H_A , \; H_{int} ] = 0 \;  .
$$
The quasi-isolation condition (7) guarantees that the matrix representation 
of all system operators from the algebra of local observables remains unchanged 
in the presence of the environment. In that sense, the latter does not disturb 
the system. Condition (7) slightly differs from the notion of a nondemolition 
measurement [14]. According to the latter, a nondemolition measurement of an 
observable, represented by the operator $\hat{A}$, is such that this operator 
is an integral of motion, $[\hat{A}, H_{AB}] = 0$. Particular systems in an 
environment satisfying Eq. (7) have been considered, assuming that this 
environment is a bath composed of an infinite number of harmonic oscillators 
[15-19]. But here, we do not specify the environment. It will be shown below 
that the quasi-isolation condition (7), as such, is sufficient for deriving 
the general properties of equilibrating quantum systems.

Let the eigenproblem for $H_A$ be of the form
\be
\label{8}
 H_A \; | \; n \; \rgl =   E_n \; | \; n \; \rgl \; .
\ee
Then, because of commutator (7), the eigenproblem for Hamiltonian (1) 
reads as
\be
\label{9}
 H_{AB} \; | \; nk \; \rgl = 
( E_n + \ep_{nk} ) \; | \; nk \; \rgl \;  ,
\ee
where $|nk \rangle \equiv |n \rangle \bigotimes |k \rangle$ and $\ep_{nk}$
is the eigenvalue of the problem
\be
\label{10}
(H_B + H_{int}) | \; nk \; \rgl = \ep_{nk} | \; nk \; \rgl \;  .
\ee

The quantities of interest are the averages of the operators $\hat{A}$ 
from the algebra of local observables acting on the space ${\cal H}_A$. 
Such averages correspond to the measurable quantities (5) describing the 
system behavior. Using the basis composed of the vectors $|nk \rangle$ 
gives
\be
\label{11}
 \lgl \; \hat A(t) \; \rgl = 
\sum_{mn} \rho_{mn}^A(t) A_{nm} \;  ,
\ee
where $A_{mn} \equiv \langle m| \hat{A} |n \rangle$. The reduced density 
matrix
\be
\label{12}
 \rho_{mn}^A(t) \equiv \lgl \; m \; |\hat{\rho}_A(t)| \; n \; \rgl  = 
\sum_k \rho_{mnk}(t)  
\ee
is the sum of the terms
\be
\label{13}
 \rho_{mnk}(t) \equiv  \lgl \; mk | \; \hat{\rho}_{AB}(t)| \; nk \; \rgl 
= \rho_{mnk}(0) \exp \{ - i (\om_{mn} + \ep_{mnk}) t \} \;  ,
\ee
with the initial values
\be
\label{14}
  \rho_{mnk}(0) \equiv 
\lgl \; mk \; | \; \hat\rho_{AB}(0) \; | \; nk \; \rgl \; ,
\ee
and with the transition frequencies
\be
\label{15}
 \om_{mn} \equiv E_m - E_n \; , \qquad 
\ep_{mnk} \equiv \ep_{mk} - \ep_{nk} \;  .
\ee

As follows from Eqs. (15),
\be
\label{16}
 \om_{nn} = 0 \; , \qquad \ep_{nnk} = 0 \;  .
\ee
Hence the diagonal elements of matrix (12) do not depend on time:
\be
\label{17}
 \rho_{nn}^A(t) = \sum_k \rho_{nnk}(0) \equiv \rho_{nn}(0) \; .
\ee
Then, separating in average (11) the diagonal and nondiagonal terms yields
\be
\label{18}
 \lgl \; \hat A(t) \; \rgl = \sum_n \rho_{nn}(0) A_{nn} +
\sum_{m\neq n} \rho_{mn}^A(t) A_{nm} \;  .
\ee

The density matrix (12) can be written as
\be
\label{19}
 \rho_{mn}^A(t) = R_{mn}(t) \exp ( - i\om_{mn} t ) \;  ,
\ee
where
\be
\label{20}
 R_{mn}(t) \equiv \sum_k \rho_{mnk}(0) \exp ( -i \ep_{mnk} t) \;  .
\ee
Introducing the density function
\be
\label{21}
g_{mn}(\ep) \equiv \sum_k \rho_{mnk}(0) \dlt(\ep-\ep_{mnk})
\ee
makes it possible to rewrite Eq. (20) as the integral transformation
\be
\label{22}
 R_{mn}(t) = 
\int_{-\infty}^{\infty} g_{mn}(\ep) e^{-i\ep t} \; d\ep \;  .
\ee

The density function (21) is normalized so that
\be
\label{23}
 \int_{-\infty}^{\infty} g_{mn}(\ep)\; d\ep = \sum_k \rho_{mnk}(0) 
\equiv \rho_{mn}(0) \; .
\ee
This allows us to represent it as
\be
\label{24}
 g_{mn}(\ep) = \rho_{mn}(0) p_{mn}(\ep) \;  ,
\ee
with the distribution function normalized to one,
\be
\label{25}
 \int_{-\infty}^{\infty} p_{mn}(\ep)\; d\ep =  1 \; .
\ee
Thence Eq. (20) can be written as
\be
\label{26}
 R_{mn}(t) = \rho_{mn}(0) D_{mn}(t) \;  ,
\ee
with the {\it decoherence factor}
\be
\label{27}
D_{mn}(t) \equiv 
\int_{-\infty}^{\infty} p_{mn}(\ep)e^{-i\ep t}\; d\ep \;  .
\ee
Consequently, matrix (19) becomes
\be
\label{28}
 \rho_{mn}^A(t) = \rho_{mn}(t) D_{mn}(t) \;  ,
\ee
where
\be
\label{29}
 \rho_{mn}(t) \equiv \rho_{mn}(0) \exp( - i\om_{mn} t ) \;  .
\ee
As a result, for average (18), we have
\be
\label{30}
 \lgl \; \hat A(t) \; \rgl = \sum_n \rho_{nn}(0) A_{nn} +
\sum_{m\neq n} \rho_{mn}(t) A_{nm} D_{mn}(t) \;  .
\ee

\section{Possible types of equilibration}

If the surrounding is finite, so that the spectrum, defined by 
eigenproblem (10), is discrete, then the density function (21) is
a sum of delta functions. Therefore factor (27) is a sum of 
exponentials and average (30) is a quasi-periodic function. In such
a case, there is no equilibration in the strict sense, as far as any
initial value of the observable will be reproduced after the 
Poincar\'{e} recurrence time. However, if the number of degrees of 
freedom, characterizing the surrounding, is sufficiently large, the 
recurrence time can be very long [12]. Then the system tends to a
quasi-equilibrium state, in which it lives during most of the time
in the temporal interval before the recurrence time.   
 
But if the surrounding is so large that can be treated as infinite,
then the multi-index $k$ becomes continuous and the sums over $k$ are
to be replaced by integrals. In that case the density function (21) 
can be measurable, similarly to the properties of densities of states 
for large statistical systems. If so, the related state distribution, 
introduced in Eq. (24), is also measurable. Then the following general 
statement holds true.

\vskip 2mm

{\bf Theorem 1}. {\it If the state distribution $p_{mn}(\varepsilon)$ is 
measurable, then the quasi-isolated quantum system equilibrates in the 
strict sense, so that}
\be
\label{31}
 \lim_{t\ra\infty} \lgl \; \hat A(t) \; \rgl  = 
\sum_n \rho_{nn}(0) A_{nn} \; .
\ee

\vskip 2mm  

{\bf Proof}. The state distribution, by assumption, is measurable and, 
by definition (25), it is $L^1$ integrable. Then, according to the 
Riemann-Lebesgue lemma [20], for integral (27), one has
$$
 \lim_{t\ra\infty} D_{mn}(t) = 0 \;  .
$$
The matrix $\rho_{mn}(t)$ depends on time through the bounded function 
$\exp(- i \omega_{mn} t)$. Since, by the Riemann-Lebesgue lemma, the 
decoherence factor tends to zero, we come to the limiting equality (31).

\vskip 2mm

As an illustration, showing how the density function (21) can become 
measurable, let us treat the index $k$ as continuous, so that function 
(21) be represented by the integral
$$
g_{mn}(\ep) = \int \rho_{mnk}(0) \dlt(\ep-\ep_{mnk} ) \;
d\mu(k) \; ,
$$
with a $d$-dimensional differential measure
$$
d\mu(k) = \frac{2\pi^{d/2}}{\Gm(d/2)} \; k^{d-1} dk \; ,
$$
corresponding to a spherically symmetric surrounding. 

The delta-function of a function $f(x)$ is defined by the expression
$$
 \dlt(f(x)) = \sum_i \; \frac{\dlt(x-x_i)}{|f'(x_i)|} \;  ,
$$
in which $x_i$ are simple zeros given by the equation $f(x_i) = 0$,
with the derivatives $f^\prime(x_i)$ being nonzero. Then, for our case,
we get 
$$
 \dlt(\ep - \ep_{mnk}) = \sum_i \; 
\frac{\dlt(k-k_i)}{|\ep_{mnk_i}'|} \; ,
$$
where $k_i = k_{imn}(\epsilon)$ is defined by the equations
$$
 \ep - \ep_{mnk_i} = 0 \; , \qquad
\ep_{mnk_i}' \equiv \frac{d\ep_{mnk_i}}{dk_i} \neq 0 \;  .
$$
This reduces the density function to the measurable form
$$
g_{mn}(\ep) = \frac{2\pi^{d/2}}{\Gm(d/2)} \sum_i
\rho_{mnk_i}(0) \; \frac{k_i^{d-1}}{|\ep_{mnk_i}'| } \; .
$$
  
The second term in Eq. (30) is caused by quantum coherence. Therefore,
its disappearance because of the action of environment, as in limit (31), 
is named the environment induced decoherence [21-23].   

The concrete expression for the decoherence factor (27) is defined by the 
state distribution $p_{mn}(\epsilon)$ induced by surrounding. Keeping in 
mind a large number of random perturbations, produced by the surrounding, 
the induced distribution, in view of the central limit theorem, can be 
modeled by the Gaussian form, with a standard deviation $\sigma_{mn}$,
$$
 p_{mn}^G(\ep) = \frac{1}{\sqrt{2\pi}\;\sgm_{mn} } \;
\exp \left ( -\; \frac{\ep^2}{2\sgm_{mn}^2 } \right ) \; .
$$ 
As a result, the decoherence factor is also Gaussian
\be
\label{32}
D_{mn}^G(t) = \exp \left \{ - \; \frac{1}{2} \; (\sgm_{mn} t)^2 
\right \} \;   .
\ee
In the case of the Lorentz distribution
$$
 p_{mn}^L(\ep) = \frac{\gm_{mn} }{\pi(\ep^2+\gm_{mn}^2) } \; ,
$$
we have the exponential form 
\be
\label{33} 
 D_{mn}^L(t) = \exp( - \gm_{mn} t) \;  .
\ee
If the induced density is of the Poisson type
$$
 p_{mn}^P(\ep) = \frac{1}{2\gm_{mn} } \; \exp \left ( -\;
\frac{|\ep|}{\gm_{mn} } \right ) \;  ,
$$
we get the decoherence factor
\be
\label{34}
 D_{mn}^P(t) = \frac{1}{1+(\gm_{mn}t)^2 } \;  .
\ee
When the surrounding is modeled by a uniform distribution
$$
p_{mn}^U(\ep) = \frac{1}{2\Dlt_{mn}} \; \Theta(\Dlt_{mn}-\ep)
\Theta(\Dlt_{mn}+\ep) \; ,
$$
the decoherence factor is of power law with oscillations:
\be
\label{35}
 D_{mn}^U(t) = \frac{\sin(\Dlt_{mn}t)}{\Dlt_{mn}t} \;  .
\ee

Under different circumstances, one can meet different behavior of the
decoherence factor. For instance, the relaxation to an equilibrium state 
can be Gaussian [24], or exponential [15,19,25], or of power law [16]. 
Generally, the influence of surrounding can be represented by a linear
combination of the above types of induced densities. Hence, the 
decoherence factor can also consist of several parts. For example, it 
can be a combination of the Gaussian and exponential terms [11,17,18].   
In any case, if the density $p_{mn}(\epsilon)$ is measurable, then a 
quasi-isolated quantum system equilibrates to a stationary state 
characterized by the diagonal elements $\rho_n \equiv \rho_{nn}(0)$, 
whose definition involves information on the initial state of the system, 
which needs to be given. The most general method of characterizing a 
stationary statistical system is by constructing the corresponding 
representative ensemble by maximizing the system entropy, under prescribed 
additional constraints [26-28], which is equivalent to the minimization 
of the information functional [29,30]. One obvious constraint is the 
normalization of $\rho_n$. Other constraints are given by the average 
values of {\it constraint operators} $\hat{C}_i$, usually corresponding 
to some observable quantities $C_i$ and initial conditions [31].  

A quasi-isolated quantum system can also thermalize, when on the manifold 
of initial conditions there exists a basin of attraction, such that limit 
(31) does not depend on the particular choice of initial conditions from 
this basin of attraction [12]. A simple case is when the system initial 
state is an eigenstate, say $|j \rangle$, so that $\rho_{nn}(0) = \delta_{nj}$, 
then limit (31) is exactly $A_{jj}$. Suppose now that the initial state is 
prepared such that $\rho_{nn}(0)$ is nonzero only for a window $\mathbb{N}_j$ 
of the indices $n \in \mathbb{N}_j$ around $j$, where the eigenvalue of $H_A$ 
in Eq. (8) is $E_j$. And let this window be so narrow that
$\Dlt A\equiv\max_{n\in\mathbb{N}_j} A_{nn} - \min_{n\in\mathbb{N}_j} A_{nn}$
be much smaller than $|A_{jj}|$. Then, by the mean value theorem, the time 
limit (31) is again $A_{jj}$ for any normalized $\rho_{nn}(0)$. Since this 
is valid for any normalized distribution, it is possible to take, for 
simplicity, the uniform expression $\rho_{nn}(0) = 1/Z_j$, where $Z_j$ is 
the number of states in $\mathbb{N}_j$. This expression corresponds to the 
microcanonical distribution, with the energy $E_j$. Since limit (31) acquires 
the microcanonical form, one can say that there happens the eigenstate 
thermalization. For isolated quantum systems, the eigenstate thermalization 
was suggested as a hypothesis by Deutsch [32] and Srednicki [33]. Here, we 
have shown that for a quasi-isolated system, with the prescribed initial 
condition, this is not a hypothesis but a rigorous asymptotic result. Note 
that the microcanonical distribution is a particular case of the general 
representative distribution [12,31]. 

It is possible, in principle, that, in addition to a measurable part of the 
density $p_{mn}(\epsilon)$, there would exist a non-measurable term of the form
$$
 p_{mn}^F(\ep) = \frac{1}{2} \sum_j c_{mnj} 
[ \dlt(\ep-\al_{mnj}) + \dlt(\ep+\al_{mnj}) ] \; ,
$$ 
with positive coefficients $c_{mnj}$ normalized so that the total density
be normalized as in Eq. (25). This would yield the appearance in factor (27) 
of the fluctuating term
\be
\label{36}
D_{mn}^F(t) = \sum_j c_{mnj} \cos(\al_{mnj} t ) \; .
\ee  
Then, instead of limit (31), we would have the asymptotic expression
\be
\label{37}
 \lgl \hat A(t) \rgl \simeq \sum_n \rho_{nn}(0) A_{nn} +
\sum_{m\neq n} \rho_{mn}(t) A_{nm} D_{mn}^F(t) \; ,
\ee
as time tends to infinity. In that case, the system does not equilibrate 
completely, but only partially, exhibiting permanent fluctuations. Though 
the temporal average of the latter expression is zero, the fluctuations  
can be rather strong and their existence may drastically change the system 
properties, hence these fluctuations cannot be neglected. A physical example 
of such a situation can be associated with heterophase fluctuations [29,34].

\section{Irreversibility of time direction}

The problem of time irreversibility has attracted great attention 
(see, e.g., Refs. [21-23,35,36]). Usually, one connects the irreversibility
with the behavior of the system thermodynamic characteristics, such as the
increase of entropy with time. The Gibbs entropy is known to be constant 
for isolated systems, because of which the Boltzmann entropy seems to be 
preferable for considering this problem [37,38]. 

Another possibility is to consider the average entropy or the average 
information. The latter is defined as the average of the information content
${\rm ln} \hat\rho_{AB}$,
\be
\label{38}
 I(t) \equiv 
{\rm Tr}_{AB} \hat\rho_{AB}(t) \ln \hat\rho_{AB} \;  ,
\ee
similarly to the definition of the operator averages (5). Then we have the 
following statement.

\vskip 2mm

{\bf Theorem 2}. {\it The average information (38) does not increase with 
time}:
\be
\label{39}
 I(t) - I(0) \leq 0 \;  .
\ee

\vskip 2mm

{\bf Proof}. It is straightforward to see that
$$
{\rm Tr}_{AB} \hat\rho_{AB}(t) \ln \hat\rho_{AB} =
{\rm Tr}_{AB} \hat\rho_{AB} \hat U^+(t) \left ( 
\ln \hat\rho_{AB}\right ) \hat U(t) =
$$
$$
 = {\rm Tr}_{AB} \hat\rho_{AB} \ln \left [ 
\hat U^+(t) \hat\rho_{AB}\hat U(t) \right ] = 
{\rm Tr}_{AB} \hat\rho_{AB} \ln \hat \rho_{AB}(-t) \;  .
$$
Therefore
$$
I(0) - I(t) = {\rm Tr}_{AB} \left [
\hat\rho_{AB} \ln \hat\rho_{AB} -
\hat\rho_{AB} \ln\hat\rho_{AB}(-t) \right ] \;   .
$$
Then we use the Gibbs-Klein inequality, proved by Gibbs [39] for classical 
distributions and generalized by Klein [40] for quantum operators. According 
to this inequality, for any two non-negative operators $\hat{A}$ and $\hat{B}$,
acting on a Hilbert space $\cal{H}$, one has
$$
 {\rm Tr}_{\cal H} \left ( \hat A \log \hat A -
\hat A \log \hat B \right ) \geq 
{\rm Tr}_{\cal H}(\hat A - \hat B) \;  .
$$
In our case, we get
\be
\label{40}
 I(0) - I(t) \geq {\rm Tr}_{AB} \left [ \hat\rho_{AB}  -
\hat\rho_{AB}(-t) \right ] \;  .
\ee
Since
$$
{\rm Tr}_{AB} \hat\rho_{AB}(-t) =  
{\rm Tr}_{AB} \hat\rho_{AB} = 1 \;  ,
$$
we come to the sought inequality (39). 

\vskip 2mm

The loss of information with time can be interpreted as the origin of the 
time irreversibility. However, the irreversibility of time is a phenomenon 
common for all macroscopic systems, whether thermodynamic or generally dynamic. 
Therefore the explanation of this phenomenon should not be based only on the 
notion of entropy. Another point of view on the time irreversibility, not 
necessarily connected with thermodynamics, relies on the fact that, strictly 
speaking, absolutely isolated systems do not exist, since any given system is 
always, at least weakly, connected to some surrounding [22,36]. A real system
can be quasi-isolated, but cannot be isolated completely, since there always 
exist small perturbations, often uncontrollable, which induce the interaction 
of the system with its surrounding [41-43]. This point of view has been 
emphasized long ago by Borel [44]. Accepting this, we come to the conclusion 
that any given real system, at least partially, equilibrates, according 
to Eqs. (31) or (37). The fact that any real system tends to an equilibrated 
state implies that there exists irreversible evolution, that is, 
{\it time is irreversible}. The partially equilibrated state can be a kind of 
a steady or quasi-steady state [45].

\section{Conclusion}

The main results and conclusions of the present paper can be formulated 
as follows: 

(i) The temporal evolution of quasi-isolated quantum systems 
can be treated on the general level, valid for systems of arbitrary 
physical nature and under any initial conditions. 

(ii) Quasi-isolated systems, with increasing time, equilibrate, at 
least partially. 

(iii) When the system equilibrates to a stationary state, the latter, 
generally, keeps information on initial conditions and is characterized 
by a representative ensemble. 

(iv) For some initial conditions, the eigenstate thermalization happens, 
being a rigorous asymptotic result. 

(v) Accepting that all real systems are at the most only quasi-isolated, 
but never absolutely isolated, defines the arrow of time, explaining why 
time is irreversible.

\vskip 5mm
The work is supported by a grant of the Russian Foundation for Basic 
Research. Discussions with E.P. Yukalova are appreciated.


\newpage


\begin{thebibliography}{99}

\bibitem{1}
E. Schr\"{o}dinger, Ann. Physik 83 (1927) 956. 

\bibitem{2}
E. Schr\"{o}dinger, Statistical Thermodynamics,
Cambridge University, Cambridge, 1952.

\bibitem{3}
J. von Neumann, Z. Phys. 57 (1929) 30. 

\bibitem{4}
J. von Neumann, Eur. Phys. J. H  35 (2010) 201.

\bibitem{5}
S.T. Smith, R. Onofrio, Eur. Phys. J. B 61 (2008) 271.

\bibitem{6}
A.V. Ponomarev, S. Denisov, P. H\"{a}nggi, Phys. Rev. Lett. 
106 (2011) 010405. 

\bibitem{7}
U. Divakaran, F. Igloi, H. Rieger, arXiv:1105.5317 (2011).

\bibitem{8}
A. Polkovnikov, Ann. Phys. (N.Y.) 325 (2010) 1790. 

\bibitem{9}
P. Reimann, New J. Phys. 12 (2010) 055027. 

\bibitem{10}
A. Polkovnikov, K. Sengupta, A. Silva, M. Vengalatore, 
Rev. Mod. Phys. 83 (2011) 863. 

\bibitem{11}
S. Yuan, J. Comput. Theor. Nanosci. 8 (2011) 889. 

\bibitem{12}
V.I. Yukalov, Laser Phys. Lett. 8 (2011) 485. 

\bibitem{13}
M.A. Cazalilla, R. Citro, T. Giamarchi, E. Orignac, M. Rigol, 
arXiv:1101.5337 (2011). 

\bibitem{14}
V.B. Braginsky, F.Y. Khalili, Rev. Mod. Phys. 68 (1996) 1.

\bibitem{15}
N.G. van Kampen, J. Stat. Phys. 78 (1995) 299. 

\bibitem{16}
J. Shao, M.L. Ge, H. Cheng, Phys. Rev. E 53 (1996) 1243. 

\bibitem{17}
D. Mozyrsky, V. Privman, J. Stat. Phys. 91 (1998) 787. 

\bibitem{18}
M. Merkli, I.M. Sigal, G.P. Berman, Ann. Phys. (N.Y.) 323 (2008) 373.

\bibitem{19}
M. Merkli, G.P. Berman, I.M. Sigal, Ann. Phys. (N.Y.) 323 (2008) 3091. 

\bibitem{20}
S. Bochner, K. Chandrasekharan, Fourier Transforms, 
Princeton University, Princeton, 1949.

\bibitem{21}
W.H. Zurek, Ann. Physik 9 (2000) 855.

\bibitem{22}
H.D. Zeh, Entropy 7 (2005) 199.

\bibitem{23}
H.D. Zeh, Found. Phys. 40 (2010) 1476. 

\bibitem{24}
D.A. Dalvit, G.P. Berman, M. Vishik, Phys. Rev. A 73 (2006) 013803.

\bibitem{25}
L.M. Duan, G.C. Guo, Phys. Rev. A 57 (1998) 737.

\bibitem{26}
C.E. Shannon, Bell Syst. Tech. J. 27 (1948) 379. 

\bibitem{27}
C.E. Shannon, Collected Papers, IEEE Press, New York, 1993.

\bibitem{28}
E.T. Janes, Phys. Rev. 106 (1957) 620.  

\bibitem{29}
V.I. Yukalov, Phys. Rep. 208 (1991) 395.  

\bibitem{30}
V.I. Yukalov, Phys. Rev. E 72 (2005) 066119.

\bibitem{31}
V.I. Yukalov, Phys. Lett. A 375 (2011) 2797.

\bibitem{32}
J.M. Deutsch, Phys. Rev. A 43 (1991) 2046. 

\bibitem{33}
M. Srednicki, Phys. Rev. E 50 (1994) 888.

\bibitem{34}
V.I. Yukalov, Int. J. Mod. Phys. B 17 (2003) 2333.

\bibitem{35}
L.S. Shulman, Time's Arrow and Quantum Mechanics, 
Cambridge University, Cambridge, 1997.

\bibitem{36}
H.D. Zeh, The Physical Basis of the Direction of Time, 
Springer, Berlin, 2001.

\bibitem{37}
J.L. Lebowitz, Physica A 194 (1993) 1.

\bibitem{38}
S. Goldstein, J.L. Lebowitz, Physica D 193 (2004) 53.

\bibitem{39}
J.W. Gibbs, Elementary Principles in Statistical Mechanics, 
Oxford University, Oxford, 1902.

\bibitem{40}
O. Klein, Z. Phys. 72 (1931) 767.

\bibitem{41}
V.I. Yukalov, Phys. Rev. E 65 (2002) 056118. 

\bibitem{42}
V.I. Yukalov, Phys. Lett. A 308 (2003) 313.

\bibitem{43}
V.I. Yukalov, Physica A 320 (2003) 149.

\bibitem{44}
E. Borel, Le Hasard, Alcan, Paris, 1924.

\bibitem{45}
D.J. Searles, L. Rondoni, D.J. Evans, J. Stat. Phys. 128 (2007) 1337. 
 
\end{thebibliography}
\end{document}